\documentstyle[aps,preprint]{revtex}

\begin{document}
\draft
\title
{\bf Disorder Effects in Superconductors with Anisotropic Pairing:\
From Cooper Pairs to Compact Bosons }
\author{ M.V.Sadovskii,\ A.I.Posazhennikova }
\address
{Institute for Electrophysics,\\ Russian Academy of Sciences,\ 
Ural Branch,\\ Ekaterinburg,\ 620049, Russia\\
E-mail:\ sadovski@ief.intec.ru} 
%\date{} 
\maketitle

%\begin{center}

{\sl Submitted to JETP Letters}

%{\sl cond-mat/ }

%\end{center}

\begin{abstract}
In the weak coupling BCS-approximation normal impurities do not
influence superconducting $T_{c}$ in significant manner in case of 
isotropic $s$-wave
pairing. However, in case of $d$-wave pairing these are
strongly pair-breaking. This fact is in rather strong contradiction with
many experiments on disordered high-$T_{c}$ superconductors assuming 
the $d$-wave nature of pairing in these
systems. With the growth of electron attraction within the Cooper pair the
system smoothly crosses over from BCS-pairs to compact Boson picture of
superconductivity. As pairing strength grows and pairs
become compact significant deviations from universal Abrikosov-Gorkov
dependence of $T_{c}$ on disorder appear in case of $d$-wave pairing with
superconducting state becoming more stable than in the weak coupling case.
As high-$T_{c}$ superconductors are actually in the intermediate region with
Cooper pairs size of the order of few interatomic lengths, these results can
explain the relative stability of $d$-wave pairing under
rather strong disordering.
\end{abstract}

\newpage
\narrowtext

It is well known that in the usual weak-coupling BCS-approximation
normal impurities do not influence superconducting $T_{c}$ in case of
isotropic $s$-wave pairing (Anderson theorem) \cite{PG}. In case of the so 
called anisotropic $s$-wave pairing $T_{c}$ reduction due to disorder is also 
relatively weak \cite{BH,FN}. However in case of $d$-wave pairing normal
impurities are strongly pair-breaking \cite{BH,FN,RD} and the universal
dependence of $T_{c}$ on disorder is expressed by the famous
Abrikosov-Gorkov equation:
\begin{equation}
ln\biggl(\frac{T_{c0}}{T_c}\biggr)=
\biggl[\Psi\biggl(\frac{1}{2}+\frac{\gamma}{2\pi T_{c}}\biggr)-
\Psi\biggl(\frac{1}{2}\biggr)\biggr] 
\label{1}
\end{equation}
where $\Psi(x)$ is digamma function, $\gamma = \pi n_{imp}v^2 N(E_{F})$ is 
the usual scattering rate of electrons, due to impurities with point-like 
potential $v$,\ which are chaotically distributed in space with some
density $n_{imp}$,\ $N(E_{F})$ - density of states at the Fermi level $E_{F}$.
From Eq.(\ref{1}) it follows directly that $T_{c}$ is completely suppressed
at some critical scattering rate $\gamma = 0.88 T_{c0}$, which determines the
appropriate critical impurity concentration or residual resistivity of the
normal state 
\begin{equation}
\rho_{AG}=\frac{2m\gamma_{c}}{ne^2}=\frac{8\pi\gamma_{c}}
{\omega_{p}^2}
%=22.1 \frac{T_{c0}}{\omega_{p}^2}
\label{2}
\end{equation}
where $n$ and $m$ are electron concentration and mass,\ $\omega_{p}$ is
plasma frequency of electrons \cite{RD}.

At present there is an emerging consensus on the $d$-wave nature of the pairing 
state in high-temperature
superconducting copper oxides \cite{H}. However the scale of the critical 
scattering rate of $\gamma_{c}\sim T_{c0}$ is in rather strong contradiction 
with the large
amount of data on disorder suppression of $T_{c}$ in these systems \cite{MS},
which apparently demonstrate superconducting state being conserved up to
disorder induced metal-insulator transition,\ i.e $\gamma\sim E_{F}\gg 
T_{c0}$. The aim of the present report is to propose some possible 
explanation of this discrepancy.

Consider the (opposite to the usual BSC-picture) limit of extremely strong
pairing interaction,\ leading to compact Boson formation \cite{NS}.
In this case $T_{c}$ is determined by the temperature of Bose
condensation of free Bosons. In case of impure system condensation point
can be determined by the following equation \cite{PP}:
\begin{equation}
\mu_{p}-\Sigma(0)=0
\label{3}
\end{equation}
where $\mu_{p}$ is the chemical potential of pairs and $\Sigma(0)$ is the
zero-frequency limit of Boson self-energy due to impurity scattering, which 
in the weak scattering approximation reduces to the one-loop expression,
corresponding to diagram shown in Fig.1:
\begin{equation}
\Sigma(\varepsilon_{n})=n_{imp}v^2\int\frac{d^3{\bf p}}{(2\pi)^3}
\frac{1}{i\varepsilon_{n}-\frac{{\bf p}^2}{2m^{\star}}+\mu_{p}}
\label{4}
\end{equation}
where $\varepsilon_{n}=2\pi nT$ is the even Matsubara frequency,\ 
$m^{\star}=2m$ is the mass of the pair,\ and we assume temperatures
$T>T_{c}$.\ In the following we consider only three-dimensional systems.
Direct calculations give:
\begin{equation}
\Sigma(0)=Re\tilde\Sigma(0)+E_{0c}
\label{5}
\end{equation}
where $E_{0c}=-\frac{m^{\star}}{\pi^2}n_{imp}v^2p_{0}$ is the band-edge 
shift due to impurity scattering \cite{MVS} ($p_{0}$ - is some cut-off in
momentum space of the order of inverse lattice spacing $a^{-1}$) and
\begin{equation}
Re\tilde\Sigma(0)=\frac{1}{\sqrt{2}\pi}n_{imp}v^{2}{m^{\star}}^{3/2}
\sqrt{|\mu_{p}|}
\label{6}
\end{equation}
Actually, $E_{0c}$ leads just to renormalization of the chemical potential:
$\tilde\mu=\mu_{p}-E_{0c}$,\ so that in renormalized form Eq.(\ref{3})
reduces to:
\begin{equation}
\tilde\mu\left(1-\frac{1}{\sqrt{2|\tilde\mu|}\pi}n_{imp}v^2{m^{\star}}^{3/2}
sign\tilde\mu\right)=0
\label{7}
\end{equation}
with the only relevant ($\tilde\mu<0$ for Bosons at $T>T_{c}$) solution of
$\tilde\mu=0$,\ i.e.\ $\mu_{p}-E_{0c}=0$,\ determining the Bose condensation
temperature of the impure system by the standard equation:
\begin{equation}
\frac{n}{2}=g\int\limits_{-\infty}^{\infty}d\varepsilon N(\varepsilon)
\frac{1}{e^{\frac{\varepsilon}{T_{c}}}-1}
\label{8}
\end{equation}
where $g=2s+1$ (for Bosons of spin $s$),\ $N(\varepsilon)$ is the impurity 
averaged density of states,\ which in case of the simplest approximation of
Eq.(\ref{4}) just reduces to $N(E-E_{0c})$ - the usual free particle
expression with energy $\varepsilon$ calculated with respect to the shifted
band-edge. Obviously we obtain the standard expression for $T_{c}$ \cite{LL}:
\begin{equation}
T_{c}=\frac{3.31}{g^{2/3}}\frac{(n/2)^{2/3}}{m^{\star}}
\label{9}
\end{equation}
which is {\em independent of disorder}. The only possible disorder effect
may be connected with exponentially small ``Lifshits tail'' in the density
of states in Eq.(\ref{8}) due to localization \cite{LGP}, which is neglected
in our simplest approximation of Eq.(\ref{4}).
Thus, our conclusion is that in case of extremely strong pairing interaction
(compact Boson picture of superconductivity) $T_{c}$ is practically disorder
independent for {\em any} value of the spin of Cooper pair, e.g.\ $s$-wave,\ 
$d$-wave etc.

It was shown rather long ago by Nozieres and Schmitt-Rink \cite{NS} for 
non impure superconductor that as the strength of the pairing interaction
grows, there is a smooth crossover of $T_{c}$ from the weak-coupling
BCS-picture to that of compact Bosons. In the impure case similar analysis
for $T_{c}$ can be performed solving the following coupled system of
equations generalizing similar equations of Ref.\cite{NS} --- the usual
equation for BCS instability: 
\begin{equation}
1-\chi(0,0)=0
\label{10}
\end{equation}
and the equation for Fermion density (chemical potential of electrons $\mu$):
\begin{equation}
\frac{1}{2}(n-n_{f})=\int\frac{d^3{\bf q}}{(2\pi)^3}\int\frac{d\omega}{\pi}
\frac{1}{exp({\frac{\omega}{T_{c}}})-1}\frac{\partial}{\partial\mu}
\delta({\bf q}\omega)
\label{11}
\end{equation}
where $n_{f}(\mu,T_{c})$ is the free Fermion part of density,
\begin{equation}
\delta({\bf q}\omega)=arctg\frac{Im\chi({\bf q}\omega)}{1-Re\chi({\bf q}
\omega)},
\label{12}
\end{equation}
and Cooper susceptibility $\chi({\bf q}\omega)$ is determined by diagrams
shown in Fig.2. In this figure the vertices contain the symmetry factors
for different types of pairing,\ e.g. in case of cubic lattice \cite{SLH}:
\begin{eqnarray}
\label{13}
\psi_{s}({\bf p})=1 \qquad\mbox{(isotropic $s$-wave)} \nonumber \\
\psi_{s'}({\bf p})=\cos p_{x}a + \cos p_{y}a + \cos p_{z}a
\qquad\mbox{(anisotropic $s$-wave)} \nonumber\\
\psi_{d_{x^2-y^2}}({\bf p})=\cos p_{x}a - \cos p_{y}a 
\qquad\mbox{($d$-wave)} \nonumber\\
\psi_{d_{3z^2-r^2}}({\bf p})=2\cos p_{z}a - \cos p_{x}a -\cos p_{y}a
\qquad\mbox{etc.}
\end{eqnarray}
Pairing interaction is assumed to have the following form:
\begin{equation}
V_{i}({\bf p},{\bf p'})=V_{{\bf pp'}}\psi_{i}({\bf p})\psi_{i}({\bf p'})
\label{14}
\end{equation}
with $\psi_{i}({\bf p})$ defined as above and pairing potential
\begin{equation}
V_{{\bf pp'}}=-\frac{V_{0}}{\sqrt{\left(1+\frac{p^2}{p_{0}^2}
\right)\left(1+\frac{p'^2}{p_{0}^2}\right)}}
\label{15}
\end{equation}
similar to that used in Ref.\cite{NS} with $p_{0}\sim a^{-1}$.

Numerical work required to solve Eqs.(\ref{10}),(\ref{11}) is
very heavy even for non impure case \cite{NS}. However, it is clear that
these equations will produce also the smooth crossover in $T_{c}$ dependence
on disorder,\ interpolating between the BCS and compact Boson limits 
discussed above. In isotropic $s$-wave case $T_{c}$ will remain practically
independent from disorder, i.e. the Anderson theorem remains valid also for
compact Boson limit. In case of $d$-wave pairing the universal dependence
of $T_{c}$ on disorder defined by Eq.(\ref{1}) ceases to be valid 
in the crossover region from large Cooper pairs to compact Bosons. The
physical reason for this is quite clear --- depairing mechanism of $T_{c}$
suppression by disorder ceases to operate with the growth of attractive
interaction within pairs, and in the strong coupling region $T_{c}$ is
determined  by Bose condensation of pairs in impure system. Qualitative
behavior of $T_{c}$ dependence on disorder is shown in Fig.3. 
It illustrates the smooth crossover in $T_{c}$ dependence on normal state
resistivity from universal Abrikosov-Gorkov dependence (curve $d$) to
$T_{c}$ independent on disorder (curve $s$). Dashed lines correspond to 
transition region and the values of coupling constant $V_{0}$ growing from
curve $1$ to curve $2$. It is clear that for $d$-wave system belonging to
this transitional region we can easily obtain superconducting state
persisting for rather large disorder with $\rho > \rho_{AG}$.

Crossover region is qualitatively defined by the simple inequality introduced
in Ref.\cite{PS}:\ $\pi^{-1} < p_{F}\xi < 2\pi$,\ where $p_{F}$ is Fermi
momentum and $\xi$ is superconducting coherence length. It appears
that high-temperature superconductors lie on the
the so-called Uemura plot \cite{U} near the ``instability'' line
$p_{F}\xi=2\pi$ \cite{PS}. 
This can explain  deviations of $T_{c}$ dependence on disorder in 
these systems from universal Abrikosov-Gorkov curve and their relative 
stability to disordering \cite{MS},\ despite the possible $d$-wave symmetry
of the pairing state.

\newpage
\begin{center}
{\bf Acknowledgements:}
\end{center}
The authors are grateful to Dr.\ A.V.Mirmelstein who actually urged us to
publish these simple conjectures.
This work was partly supported by the grant 96-02-16065 of the Russian 
Foundation for Basic Research, as well as by the grant IX.1 of the State 
Program ``Statistical Physics''.

\newpage
\begin{center}
{\bf Figure Captions:} 
\end{center}

Fig.1. One-loop Boson self-energy due to random impurity scattering.

Fig.2. (a) Diagram representation of Cooper susceptibility $\chi({\bf q}\omega)$.
$V$ --- pairing potential.
$\Gamma$ --- impurity scattering vertex-part in Cooper channel, defined by
the ``ladder''approximation (b).

Fig.3. Qualitative dependence of transition temperature $T_{c}$ on disorder 
(normal state residual resistivity $\rho$).\ Curve $d$ --- universal
Abrikosov-Gorkov dependence of Eq.(\ref{1}).\ Curve $s$ --- the case of
isotropic $s$-wave pairing.\ Dashed curves --- $d$-wave pairing in
crossover region from BCS pairs to compact Bosons.

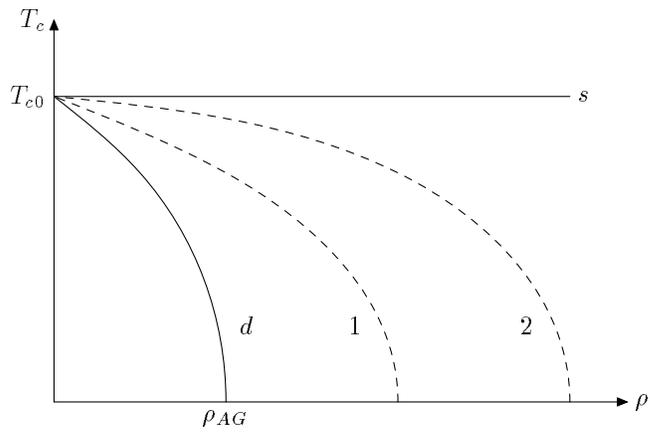
\end{document}